\documentclass[a4paper,11pt]{article}
\usepackage{pos}
\usepackage{tikz}
\usepackage{tikz-feynman}
\usepackage{subcaption}
\usepackage{adjustbox}

\title{Measurements of electroweak penguin and
lepton-flavour violating $B$ decays to final states
with missing energy at Belle and Belle II}

\author*[a,b]{Meihong Liu}

\onbehalf{on behalf of the Belle and Belle II collaborations}

\affiliation[a]{Jilin University,\\
Changchun 130012, People’s Republic of China}
\affiliation[b]{Deutsches Elektronen–Synchrotron,\\
Hamburg 22607, Germany}

\emailAdd{liumeihong@jlu.edu.cn}

\abstract{

The Belle and Belle~II experiments have accumulated a data set of $1.2~\mathrm{ab}^{-1}$ of $e^+e^- \to B\bar{B}$ collisions at the $\Upsilon(4S)$ resonance. Owing to the clean event environment and well-constrained initial-state kinematics, these data are ideally suited for searches for rare electroweak–penguin and lepton-flavour-violating $B$ decays with missing energy from neutrinos. We report results on $b\to s\nu\bar{\nu}$ processes and the interpretation, together with searches for $B\to K^{*0}\tau^+\tau^-$ and for the LFV decays $B^0\to K_S^0\tau^\pm\ell^\mp$ and $B^0\to K^{*0}\tau^\pm\ell^\mp$ ($\ell=e,\mu$).

}  

\FullConference{%
XIII International Conference on Kaon Physics (Kaon 2025), \\
  8-12 September 2025\\
  Mainz, Germany
}

\begin{document}
\maketitle

\section{Introduction}

\begin{figure}[htp]
    \centering
    \begin{subfigure}[b]{0.25\linewidth}
        \centering
        \includegraphics[width=\linewidth]{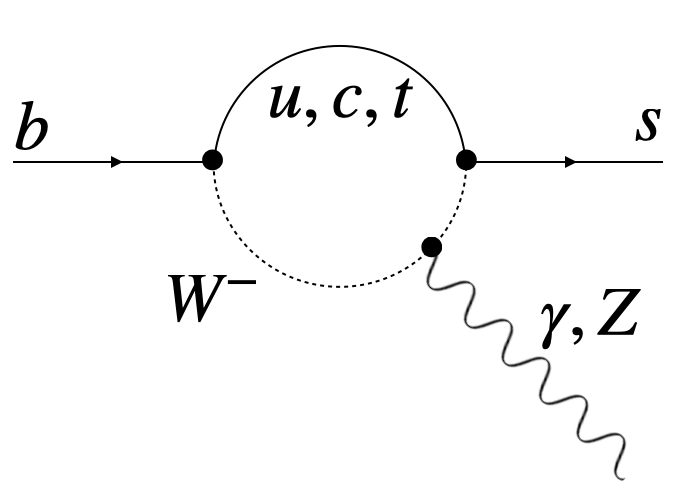}
        \caption{SM contributions}
        \label{fig:diagram1}
    \end{subfigure}
    \hfill
    \begin{subfigure}[b]{0.25\linewidth}
        \centering
        \includegraphics[width=\linewidth]{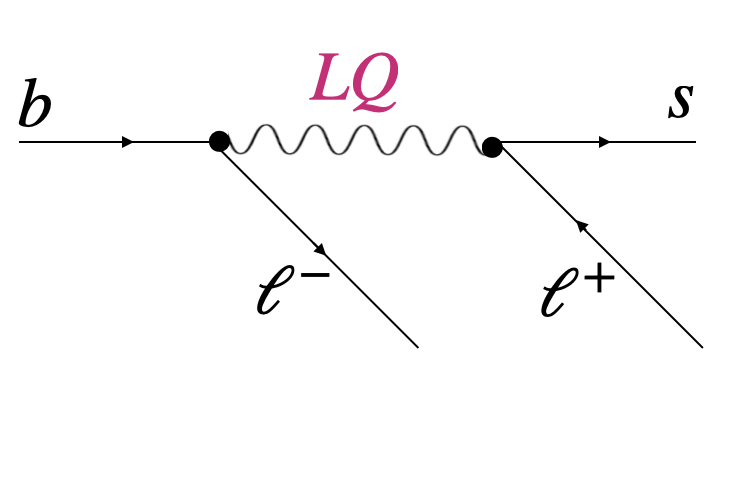}
        \caption{Leptoquark scenario}
        \label{fig:diagram2}
    \end{subfigure}
    \hfill
    \begin{subfigure}[b]{0.25\linewidth}
        \centering
        \includegraphics[width=\linewidth]{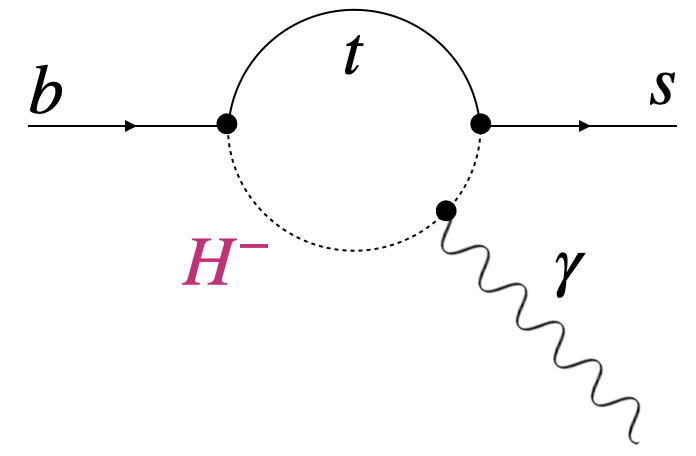}
        \caption{Charged Higgs scenario}
        \label{fig:diagram3}
    \end{subfigure}
    \caption{Diagrams representing $b\to s$ transitions: SM contributions (a) and possible NP scenarios (b,c).}
    \label{fig:diagram}
\end{figure}

The quark transition $b\to s$, mediated by a flavour-changing neutral current (FCNC), is a sensitive probe of physics beyond the Standard Model (SM). In the SM, FCNC processes are highly suppressed at tree level and occur only through loop diagrams such as the electroweak--penguin amplitudes in Fig.~\ref{fig:diagram1}, leading to small branching fractions of $\mathcal{O}(10^{-7})$--$\mathcal{O}(10^{-5})$. Searches at Belle~(II) aim to reveal possible New-Physics (NP) effects via modified rates, either from new tree-level interactions (Fig.~\ref{fig:diagram2}) or reduced GIM suppression in loops (Fig.~\ref{fig:diagram3}). Additionally, NP mediators might cause the decay rates of lepton-flavour violating (LFV) decays, which are forbidden in the SM, to become non-zero.

Belle~\cite{bdetector} operated at the KEKB asymmetric $e^+e^-$ collider~\cite{kekb} in Tsukuba, Japan. Its successor, Belle~II~\cite{b2detector}, runs at SuperKEKB, designed to reach about forty times higher instantaneous luminosity~\cite{superkekb}. (Super)KEKB collides $e^+e^-$ at the $\Upsilon(4S)$ resonance, producing almost exclusively $B\bar B$ pairs in a low-background environment compared to hadron colliders. The data samples used here correspond to 711\,fb$^{-1}$ from Belle and 365\,fb$^{-1}$ from Belle~II. Since $\Upsilon(4S)\to B\bar B$ produces no additional particles, missing energy from neutrinos in the signal decay can be inferred from the accompanying $B$ meson ($B_{\rm tag}$). The large solid-angle coverage and well-constrained initial state of Belle~(II) make it particularly well suited for analyses with missing energy, using the two tagging methods described below.

\noindent\textbf{Inclusive tagging}: this method exploits the inclusive properties of the accompanying $B$ decay. It offers high efficiency, but suffers from large background and a strong reliance on simulation.

\noindent\textbf{Hadronic tagging}: fully reconstructs specific hadronic $B$ decays using the machine-learning-based Full Event Interpretation (FEI) algorithm~\cite{fei}. This approach greatly suppresses combinatorial background and provides the strongest kinematic constraint on $B_{\rm sig}$ through ${\bf p}(B_{\rm sig})=-{\bf p}(B_{\rm tag})$ in the $\Upsilon(4S)$ rest frame, but its efficiency is typically below 1\%.

\smallskip
This report is organized as follows. Sections~\ref{KNN}, \ref{SNN}, and~\ref{KTT} present the reinterpretation of the FCNC decay $B^+\!\to K^+\nu\bar\nu$~\cite{knn_reint}, and the Belle~II searches for inclusive $B\to X_s\nu\bar\nu$~\cite{svv} and $B^0\!\to K^{*0}\tau^+\tau^-$~\cite{ktt}. Section~\ref{KSTL} reports the search for the LFV decay $B^0\!\to K_S^0\tau^{\pm}\ell^{\mp}$~\cite{kstauell} and $B^0\!\to K^{*0}\tau^{\pm}\ell^{\mp}$~\cite{kstartauell} $(\ell=e,\mu)$ using Belle and Belle~II data.

\section{$B^+\to K^+\nu\bar\nu$ reinterpretation}
\label{KNN}

We recently determined the branching fraction of the $B^{+}\to K^{+}\nu\bar{\nu}$ decay based on an integrated luminosity of 362~fb$^{-1}$ collected on the $\Upsilon(4S)$ resonance data under the assumption of Standard Model kinematics, providing the first evidence fo this decay using combined inclusive and hadronic $B$-tagging method~\cite{knn}. 
In this analysis, backgrounds are suppressed using two consecutive Boosted Decision Trees (BDT$1$ and BDT$2$). The signal yield is extracted using a two-dimensional binning scheme in the transformed BDT$2$ output, $\eta_{\mathrm{BDT_2}}$, and the reconstructed momentum-transfer squared, $q^2_{\mathrm{rec}}$, which is divided into $4\times 3$ bins.

\begin{figure}[h]
    \centering
    \begin{subfigure}[b]{0.48\linewidth}
        \centering
        \adjustbox{valign=b}{\includegraphics[width=\linewidth]{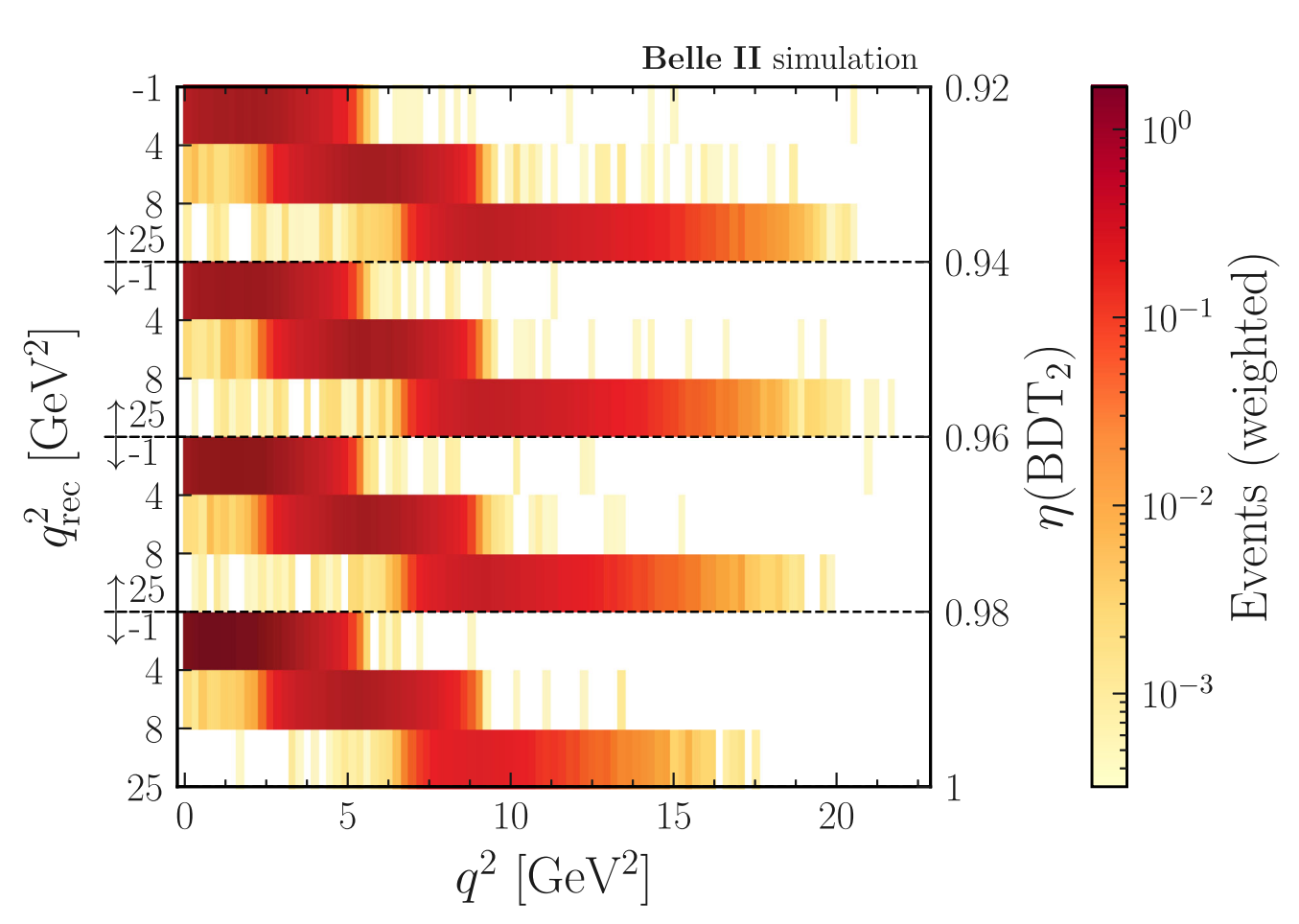}}
        \caption{}
        \label{fig:knn1}
    \end{subfigure}
    \hfill
    \begin{subfigure}[b]{0.45\linewidth}
        \centering
        \adjustbox{valign=b}{\includegraphics[width=\linewidth]{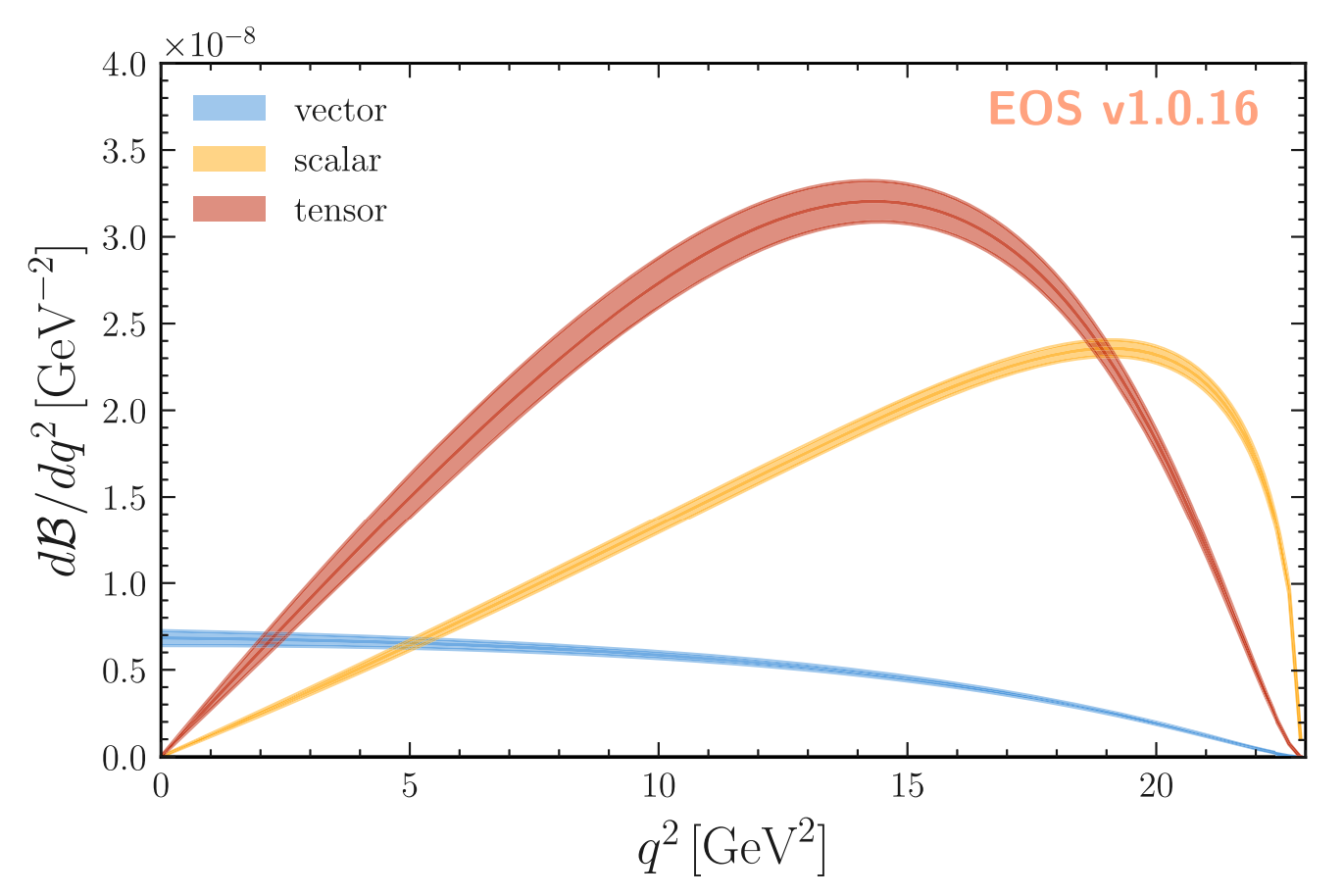}}
        \caption{}
        \label{fig:knn2}
    \end{subfigure}

    \caption{(a) ITA binned joint number densities. (b) $B^+\to K^+\nu\bar{\nu}$ differential branching fraction prediction.
}
    \label{fig:knn_inc}
\end{figure}

The reinterpretation strategy is straightforward~\cite{reint}. Instead of repeating the full analysis for every new physics model, we begin with simulated SM signal events and construct a model-agnostic number density, $n_0(x)$, where $x$ denotes the fit observables, namely $\eta_{\mathrm{BDT_2}}$ and $q^2_{\rm rec}$ as shown in Eq.~\ref{eq:1}. These variables form the basis of the fit used in the $B^{+}\!\to K^{+}\nu\bar{\nu}$ analysis. The expected event yield in a given point $x$ is obtained from the selection efficiency and the model prediction. For a new model, we simply reweight the SM events in each $q^2$ bin by the ratio of the new-model prediction to the SM one, $\frac{\sigma_1}{\sigma_0}$, which provides the predicted distribution for that model. The heatmap illustrates the weighted signal events, where the horizontal axis corresponds to the generated $q^2$ and the vertical axis shows the binning used in the $B^{+}\!\to K^{+}\nu\bar{\nu}$ analysis as shown in Fig.~\ref{fig:knn1}.

\begin{equation}\label{eq:1}
n_{1}(x)=\sum_{\text{$q^{2}$ bins}}
n_{0,q^{2}}(x)\left[\frac{\sigma_{1}(q^{2})}{\sigma_{0}(q^{2})}\right],
\qquad x\equiv \left(q^{2}_{\mathrm{rec}},\, \eta_{\mathrm{BDT_2}}\right).
\end{equation}

The Weak Effective Theory (WET) provides a unified framework for studying possible NP effects, as it describes contributions from both the SM and NP through effective operators. In this formalism, vector, scalar, and tensor Wilson coefficients contribute to the differential branching fraction of the $B^{+}\!\to K^{+}\nu\bar{\nu}$ decay as predicted by the WET is given by~\cite{wet1,wet2} 
and the predicted kinematic distributions of the respective vector, scalar and tensor operators are shown in Fig.~\ref{fig:knn2}.


\begin{figure}[h]
    \centering
    \begin{subfigure}[b]{0.42\linewidth}
        \centering
        \adjustbox{valign=b}{\includegraphics[width=\linewidth]{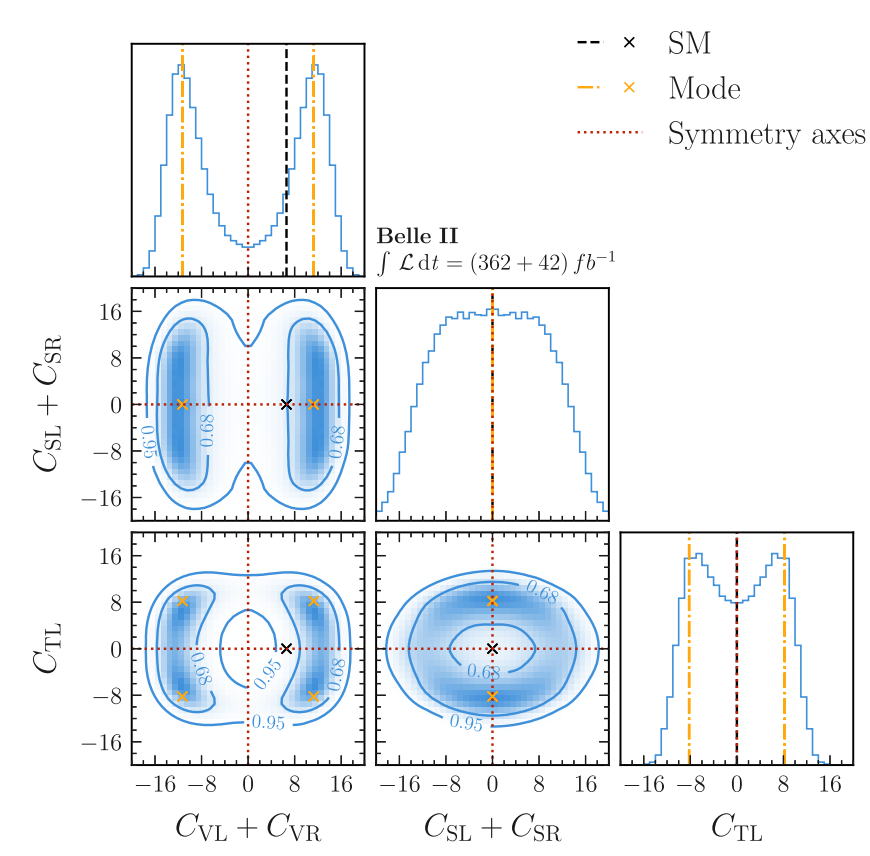}}
        \caption{}
        \label{fig:knn3}
    \end{subfigure}
    \hfill
    \begin{subfigure}[b]{0.42\linewidth}
        \centering
        \adjustbox{valign=b}{\includegraphics[width=\linewidth]{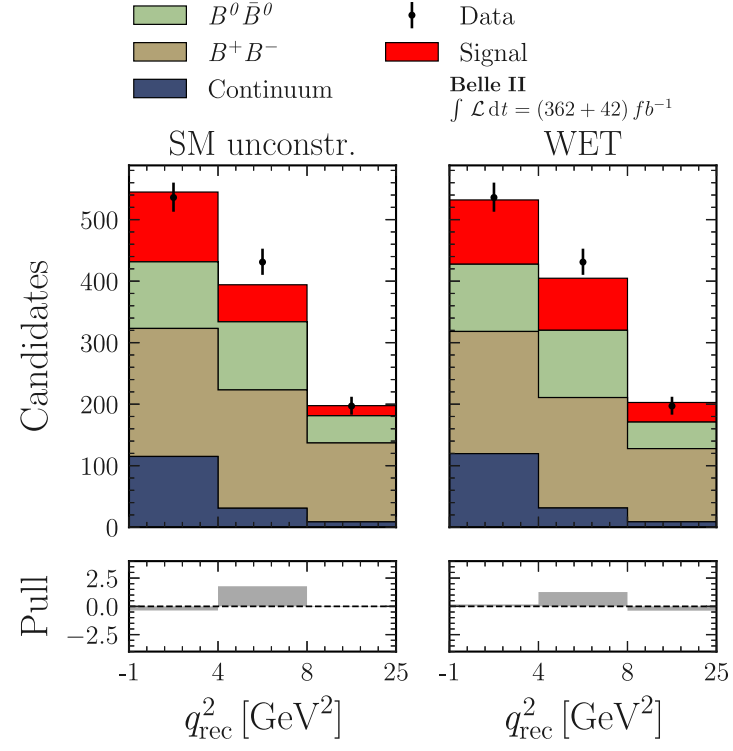}}
        \caption{}
        \label{fig:knn4}
    \end{subfigure}
    \caption{(a) The marginalized posterior for the Wilson coefficients. Dashed black lines indicate the SM predictions, and the yellow lines correspond to the posterior modes. (b) Observed and predicted best-fit yields in the highest sensitivity bins.
}
    \label{fig:knn_inc}
\end{figure}

We extract the marginal posterior distributions for the unconstrained combinations $C_{VL}+C_{VR}$, $C_{SL}+C_{SR}$, and $C_{T}$, which are treated as the parameters of interest together with the relevant hadronic form factors. The posterior maximum is located at $(11.3,\;0.0,\;8.2)$ in this parameter space. The corresponding 95\% credible intervals are $[1.9,\,16.2]$, $[0.0,\,15.4]$, and $[0.0,\,11.2]$, respectively. The diagonal panels of the figure display the one-dimensional posterior densities, while the off-diagonal panels show two-dimensional sample densities (Fig.~\ref{fig:knn3}).

Our work obtains that an enhanced vector contribution, together with a 
non-zero tensor one, provide a better description of data. As shown in Fig.~\ref{fig:knn4}~(b), the fit performed with WET-weighted events shows an improved agreement with the data compared to the SM-only unconstrained fit. Finally, we perform a fit to the $q^{2}$ spectrum using the WET-weighted events and obtain a significance of 3.3$\sigma$ for the observation of the $B^+\to K^+\nu\bar{\nu}$ signal.

\section{$B \to X_s \nu\bar\nu$ search at Belle~II using hadronic $B$-tagging at Belle II}
\label{SNN}
The second FCNC analysis presented focuses on the inclusive decay $B \to X_s \nu\bar{\nu}$ searches at Belle II. In SM, its branching fraction is predicted to be $2.9 \times 10^{-5}$ with high precision~\cite{svv_sm}. This channel is highly sensitive to several potential sources of NP and provides a complementary probe to exclusive $B\to K\nu\bar{\nu}$ searches. ALEPH reported the UL at $10^{-4}$ level over two decades ago~\cite{svv_aleph}.

In this study, a hadronic $B$-tagging algorithm is applied on the $B_{\rm tag}$ side. On the signal side, $X_s$ is reconstructed in a broad set of final states containing a kaon, including $K n\pi$ ($n=0,1,2,3,4$), $3K$, and $3K\pi$ modes. To ensure the robustness of the analysis, multiple control samples are employed to validate the reconstruction and background suppression strategies. A BDT is used for background rejection, and its efficiency is calibrated using a $B \to X_s J/\psi(\to\mu^+\mu^-)$ control sample. In this calibration, the $J/\psi$ is removed and the $X_s$ kinematics are reweighted to simulate the $X_s \nu\bar{\nu}$ signal in both data and simulation.

Additional normalization of background components is performed using off-resonance data for the $q\bar{q}$ continuum, while the BDT output and $M_{\rm bc}$ sidebands are employed to determine the normalization factor for the $B\bar{B}$ background. Systematic uncertainties are evaluated from the simulation sample size, major $B$ meson decay fractions, and signal MC generation. 

\begin{figure}[h]
    \centering 
 \includegraphics[width=0.5\linewidth]{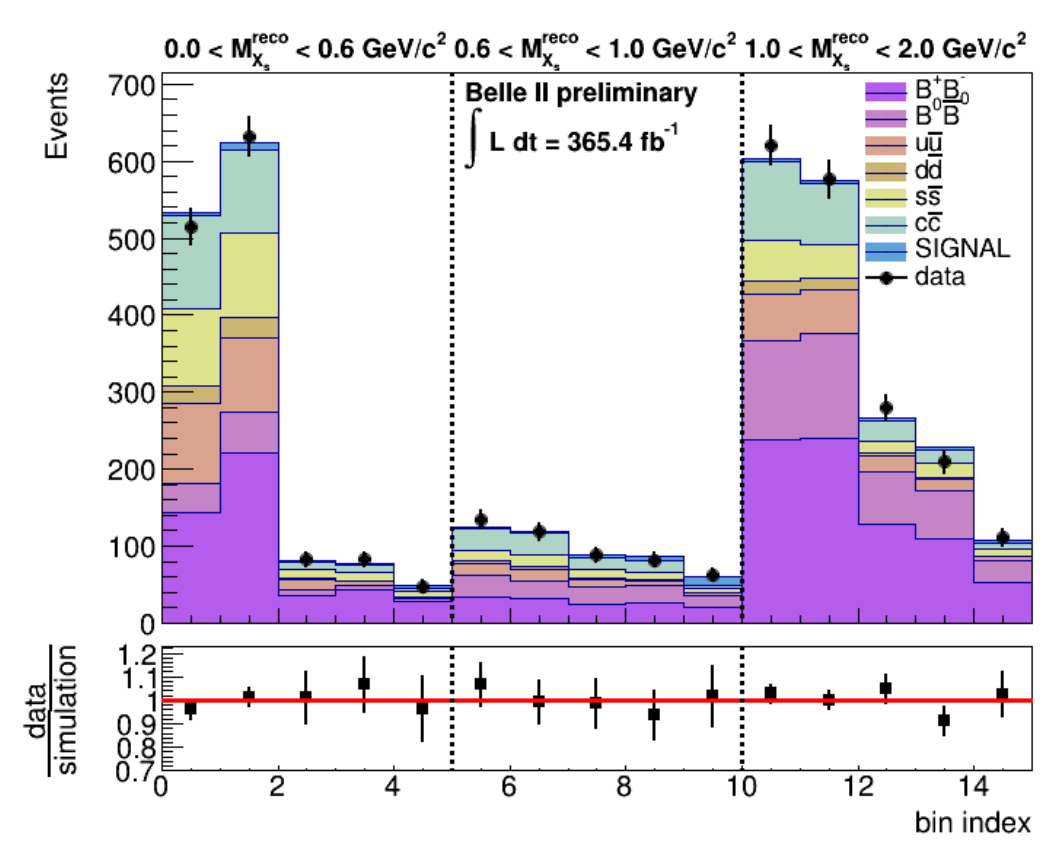}
    \caption{The bin index distribution after the fit, for data and histogram templates.}
    \label{fig:snn}
\end{figure}

A two-dimensional fit is performed using the BDT output $O'$ and the reconstructed $X_s$ mass as shown in Fig.~\ref{fig:snn}. The mass spectrum is divided into three regions: Region~1, enriched in $K$; Region~2, enriched in $K^{*0}$; and Region~3, corresponding to the non-resonant $K n\pi$ modes. 

Table~\ref{tab:efficiency_results} summarizes the selection efficiencies and fit results for the three mass regions, together with the corresponding upper limits (ULs). The UL in the $K^+$-enhanced region is consistent with the hadronic $B$-tagging Belle~II measurement of $B^+ \to K^+ \nu\bar{\nu}$. Combining the ULs from all three regions yields an overall upper limit of $3.6 \times 10^{-4}$, which constitutes the most stringent constraint on the inclusive $B \to X_s \nu\bar{\nu}$ UL to date.

\begin{table}[h!]
\centering
    \caption{Efficiencies ($\epsilon$), signal yields ($N_{\rm sig}$), and
branching fraction ($\mathcal{B}$) central values and upper limits. }
\begin{tabular}{c c c c c c c}
\hline\hline
$M_{X_s}\ [\mathrm{GeV}/c^2]$ & $\epsilon\ [10^{-3}]$ & $N_{\rm sig}$ & \multicolumn{3}{c}{$\mathcal{B}\ [10^{-5}]$} \\
\cline{4-6}
 & & & central value & UL$_{\rm obs}$ & UL$_{\rm exp}$ \\
\hline
$[0, 0.6]$ & 2.93 & $6^{+18+19}_{-17-16}$ & $0.3 \pm 0.8^{+0.9}_{-0.7}$ & 2.2 & 2.0 \\
$[0.6, 1.0]$ & 1.32 & $36^{+27+31}_{-26-26}$ & $3.5^{+2.6+3.1}_{-2.5-2.6}$ & 9.5 & 6.6 \\
$[1.0, m_B]$ & 0.62 & $24^{+44+62}_{-43-53}$ & $5.1^{+9.2+12.9}_{-8.8-11.0}$ & 31.2 & 26.7 \\
Full range & 0.97 & $66^{+64+95}_{-62-81}$ & $8.8^{+8.5+12.6}_{-8.2-10.8}$ & 32.2 & 24.4 \\
\hline\hline
\end{tabular}
\label{tab:efficiency_results}
\end{table}

\section{Search for $B^0\to K^{*0}\tau^+\tau^-$ using hadronic $B$-tagging at Belle~II}\label{KTT}

Searches for $b \!\to\! s\,\tau^{+}\tau^{-}$ have been performed at LHCb, BaBar and Belle~\cite{ktt_lhcb,ktt_babar,ktt_belle}, but no signal has been seen, mainly due to the difficult reconstruction of $\tau$ decays and the tiny SM branching fractions of $\mathcal{O}(10^{-7})$~\cite{ktt_sm1,ktt_sm2}. Belle set an upper limit of $3.1\times10^{-3}$ at 90\%~CL for $\mathcal{B}(B^0\!\to K^{*0}\tau^{+}\tau^{-})$~\cite{ktt_belle}. NP scenarios that also explain the $R(D^{(*)})$ anomaly can enhance these rates~\cite{rd}. 

Belle~II searches for $B^0\!\to K^{*0}\tau^{+}\tau^{-}$ using hadronic $B$-tagging with the FEI algorithm and additional $\tau$ decay modes to increase efficiency. Based on the $\tau$ decay products ($t_\tau$), events are classified into $\ell\ell$, $\ell\pi$, $\pi\pi$, and $\rho X$ categories, with $\ell\ell$ being the most sensitive. A BDT is trained to suppress background using missing-energy variables, unreconstructed calorimeter energy, the $K^{*0}t_\tau$ mass, and $q^2$.

\begin{figure}[htbp]
    \centering 
    \includegraphics[width=0.95\linewidth]{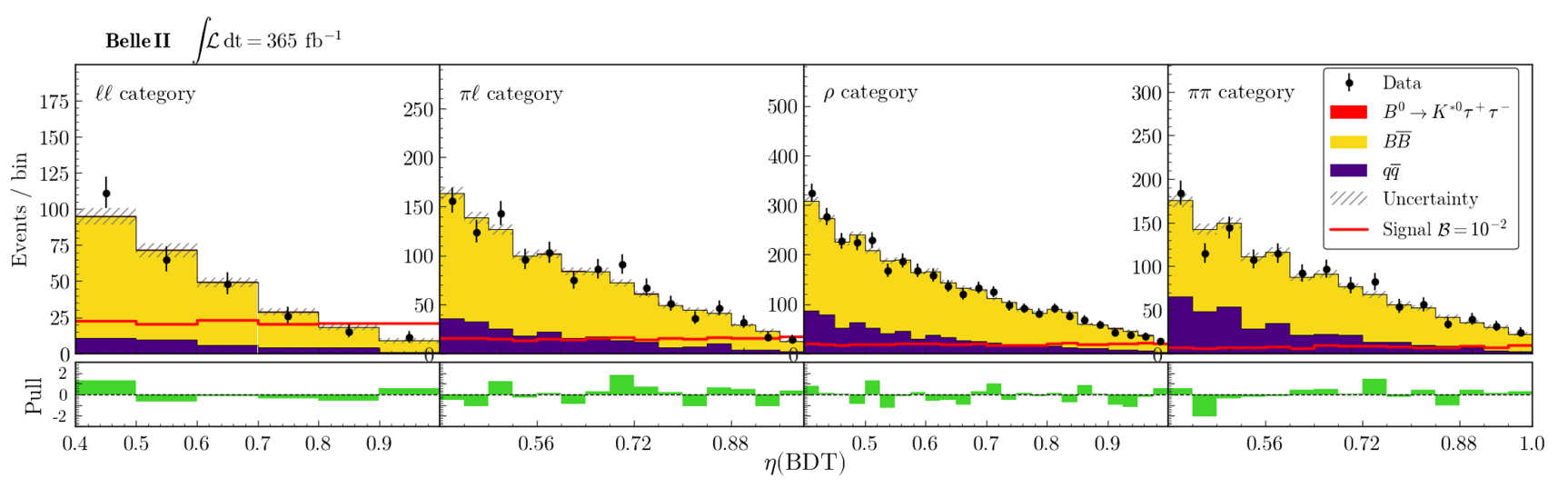}
    \caption{
    Distributions of $\eta$~(BDT) in $\text{SR}$ for the four signal categories. 
    The $B^0 \rightarrow K^{*0} \tau^+ \tau^-$ signal, fitted with a branching fraction of $[-0.15 \pm 1.01] \times 10^{-3}$ and scaled assuming a branching fraction of $10^{-2}$, is shown as a reference. 
    The bottom panel shows the pull distributions.
    }
    \label{fig:kstartautau}
\end{figure}

A simultaneous fit to the transformed BDT outputs (Fig.~\ref{fig:kstartautau}) shows no significant signal. We obtain a limit of $\mathcal{B}^{\mathrm{UL}} = 1.8\times10^{-3}$ at 90\% CL, which improves the Belle constraint by approximately a factor of two while using only half of its data. This enhancement arises from a more powerful $B$-tagging algorithm, the application of multivariate methods, and the inclusion of extra final states. This is the most stringent limit on $B^0\!\to K^{*0}\tau^{+}\tau^{-}$ and on $b\!\to\! s\,\tau^{+}\tau^{-}$ transitions to date.

\section{Search for $B^0\to K_S^{0}\tau^\pm\ell^\mp$ and $B^0\to K^{*0}\tau^\pm\ell^\mp$, $(\ell=e,\mu)$  using hadronic $B$-tagging at Belle and Belle~II}
\label{KSTL}


The Belle $\Upsilon(4S)$ dataset, the largest collected at $B$ factories, remains invaluable for searches of rare or forbidden $B$ decays. Converting Belle data into a \texttt{basf2}\footnote{Belle II analysis software framework~\cite{basf2}}-compatible format allows these searches to leverage Belle II analysis tools~\cite{b2bii}. Motivated by persistent anomalies in $B$ decays, extensions of the SM~\cite{lfvinb2s,b2sll,lfvinb} predict enhancements of LFV $b \rightarrow s \tau \mu$ decays, strictly forbidden in the SM, up to $\mathcal{O}(10^{-6})$.

We present recent results from Belle and Belle II on the LFV decays $B^0 \rightarrow K_S^0 \tau \ell$ and $B^0 \rightarrow K^{*0} \tau \ell$. The analysis strategy for these two channels is similar. We combine Belle and Belle II datasets to maximize statistics and use hadronic $B$-tagging to fully reconstruct the four-momentum of the tag-side $B$ meson. Unlike decays such as $B \rightarrow K \nu \bar{\nu}$ or $B \rightarrow K \tau \tau$, in $B \rightarrow K \tau \ell$ all neutrinos originate from the same $\tau$. This allows the 4-momentum of the 
$\tau$ to be derived using energy/momentum conservation and the knowledge 
on the collision kinematics, along with that of the reconstructed $B_{\rm tag}$, 
$K$ meson and $e/\mu$ lepton. The signal exhibits excellent resolution in the recoiling mass, while background is flat without peaking structures.  

The dominant background arises from semileptonic $B \rightarrow D$ decays, where the $K$ and $\tau$ decay product originate from the same $D$ meson. This background is visible as a threshold in the invariant mass distribution of the $K$ and $\tau$ daughter. We suppress it by requiring events above the $D$ mass threshold or vetoing the $D$ mass peak, and then employ a BDT to reduce the remaining background.  

By combining Belle and Belle II datasets, we performed the first search for $B^0 \rightarrow K_S^0 \tau \ell$ and set the first limit on this channel, reaching the $10^{-6}$ level ($[0.8- 3.6]\times 10^{-5}$). A simultaneous fit to Belle and Belle II datasets was used to derive the upper limit for $B^0 \rightarrow K^{*0} \tau \ell$ at the $10^{-5}$ level ($[2.9- 6.4]\times 10^{-5}$), consistent with previous results from LHCb~\cite{kstartauell_lhcb}.

\begin{figure}[h]
    \centering    \includegraphics[width=0.32\linewidth]{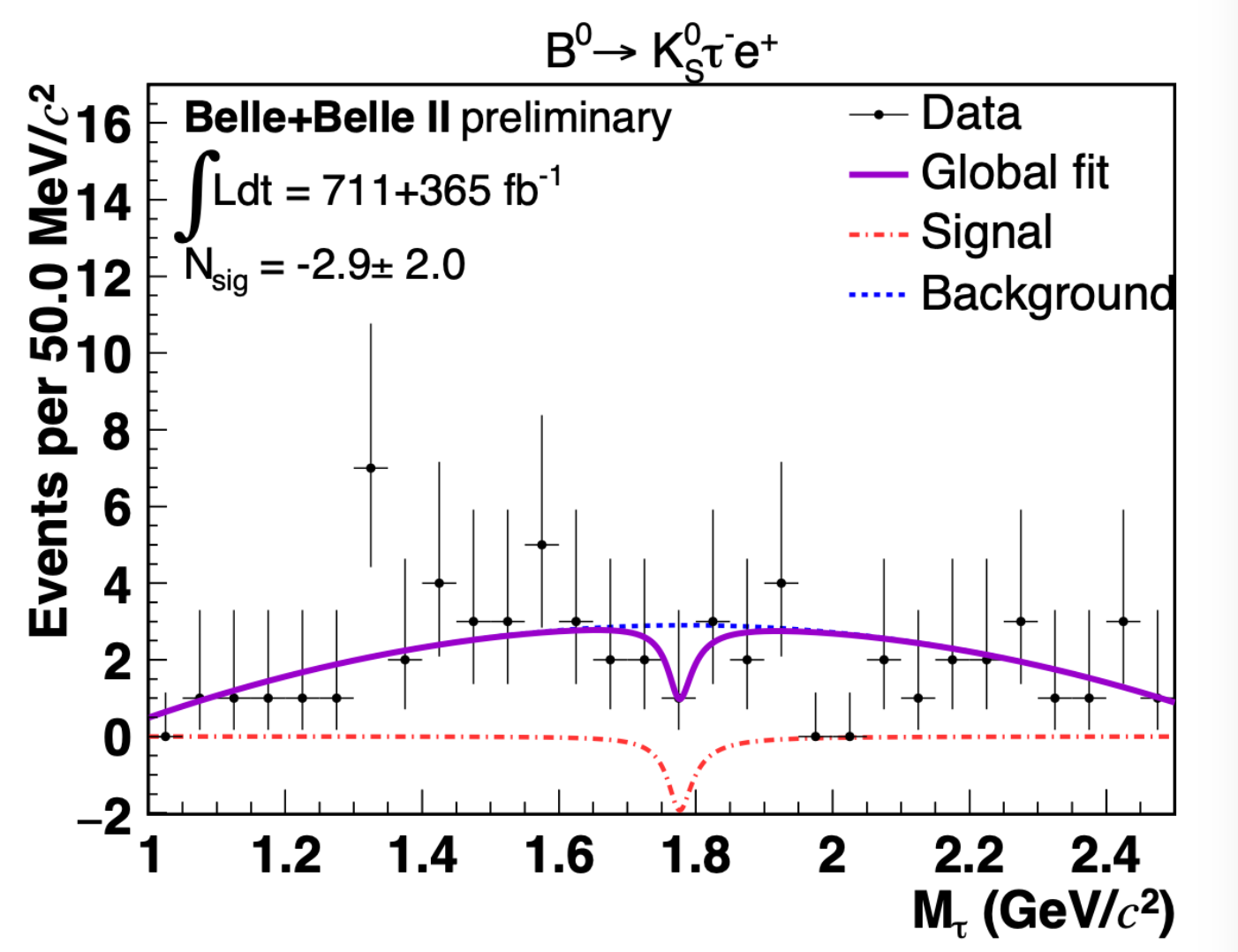}    \includegraphics[width=0.67\linewidth]{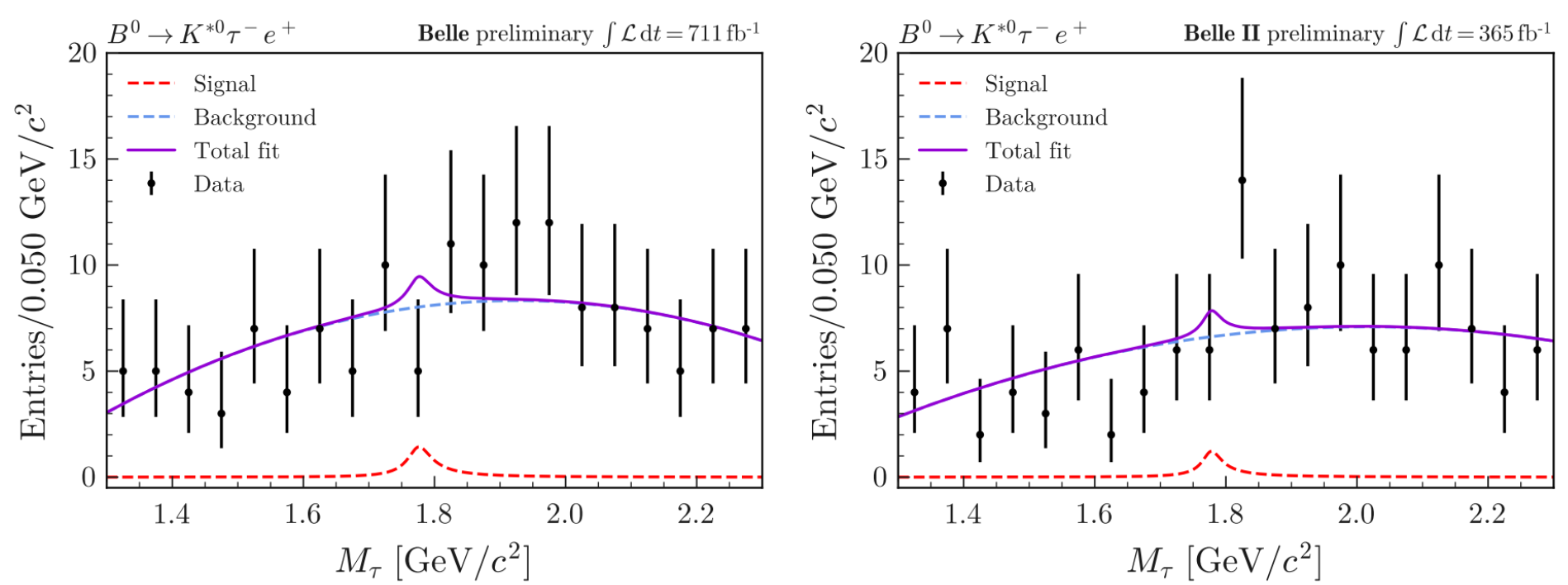}    
    \caption{Observed $M_\tau$ fits ($B^0 \rightarrow K_S^0(K^{*0}) \tau^- e^+$ as an example). The black dots with error bars show the data, the red curve shows the signal component, the blue curve shows the background component, and the purple curve shows the global fit.}
    \label{fig:kstauell}
\end{figure}

\section{Conclusion}

The Belle II detector, with its solid-angle coverage and well-known initial kinematics, offers an ideal place for studying $B$ meson decays involving missing energy through various tagging methods. Current results show that Belle II can achieve competitive, or even world-leading, precision already with a relatively limited amount of data. 


We provide a model-agnostic reinterpretation of the $B^{+}\!\to K^{+}\nu\bar{\nu}$ result~\cite{knn} within the WET framework and reports the first credible intervals for the $b\!\to s$ Wilson coefficients derived from Belle~II data. By publishing the model-agnostic likelihood, this study enables statistically rigorous tests of alternative new-physics scenarios and sets a template for future Belle~II reinterpretations.

Additionally, the FCNC processes $B \to X_s \nu\bar{\nu}$ and $B^0 \to K^{*0}\tau^+\tau^-$ have been studied using hadronic $B$-tagging with the same Belle~II data. The most stringent upper limits for these channels are determined to be $3.6 \times 10^{-4}$ and $1.8 \times 10^{-3}$ at 90\%\,CL, respectively. These results constitute the strongest constraints to date on the inclusive $B \to X_s \nu\bar{\nu}$ decay as well as on $b \to s\tau^+\tau^-$ transition.

The LFV searches $B^0\to K_S^0(K^{*0})\,\tau^\pm\ell^\mp$ ($\ell=e,\mu$) are performed using hadronic $B$-tagging with the combined Belle and Belle~II samples. The first limits on $B^0\to K_S^0\tau^\pm\ell^\mp$ are set at $[0.8\!-\!3.6]\times10^{-5}$ (90\%\,CL), providing the strongest or among the strongest constraints on $b\to s\tau\ell$ transitions. For $B^0\to K^{*0}\tau^\pm\ell^\mp$, the obtained limits of $[2.9\!-\!6.4]\times10^{-5}$ (90\%\,CL) are consistent with LHCb measurements.



\end{document}